# GeoCMS : TOWARDS A GEO-TAGGED MEDIA MANAGEMENT SYSTEM

Jang You Park[1,+], YongHee Jung[2,+], Wei Ding[1], Kwang Woo Nam[1,*]

[1] School of Computer, Information, and Communication Engineering, Kunsan National University,
54150, Gunsan, Jeonbuk, Republic of Korea - (dingwindivid, parkstar823)@gmail.com, kwnam@kunsan.ac.kr
[2] Turbosoft Inc., 604 Chungbuk Venture Center, 194-41 Osong-saengmyeong-1-ro
28160, Cheongju, Chungbuk, Republic of Korea - jyh2380@naver.com

Commission IV, WG IV/4

**ABSTRACT:**

In this paper, we propose the design and implementation of the new geo-tagged media management system. A large amount of daily geo-tagged media data generated by user's smart phone, mobile device, dash cam and camera. Geo-tagged media, such as geovideos and geophotos, can be captured with spatial temporal information such as time, location, visible area, camera direction, moving direction and visible distance information. Due to the increase in geo-tagged multimedia data, the researches for efficient managing and mining geo-tagged multimedia are newly expected to be a new area in database and data mining. This paper proposes a geo-tagged media management system, so called Open GeoCMS(Geo-tagged media Contents Management System). Open GeoCMS is a new framework to manage geo-tagged media data on the web. Our framework supports various types which are for moving point, moving photo - a sequence of photos by a drone, moving double and moving video. Also, GeoCMS has the label viewer and editor system for photos and videos. The Open GeoCMS have been developed as an open source system.

## 1. INTRODUCTION

A large amount of daily geo-tagged media data generated by user's smart phone, mobile device, dash cam and camera. Geo-tagged media captured with spatial temporal information such as time, location, visible area, camera direction, moving direction and visible distance information. Due to the increase in geo-tagged multimedia data, the researches for efficient managing and mining geo-tagged multimedia are newly expected to be a new area in database and data mining (Ding et al., 2019 and Ferreira et al., 2013).

Geo-tagged media object contains typically the following data types.
- GPS trajectories: Most of smart devices have GPS for the map and location-based services. Many apps will store user's or vehicle's trajectories to provide services.
- Geo-tagged photos: When a photo was taken in smart phone, it will store the device's location and view direction of the camera.
- Geo-tagged videos: As like geo-tagged photos, videos include gps trjaectories. Furthermore, dash cams of vehicle have accelerometer values to detect dangerous situation.

This paper proposes a geo-tagged media management system, so called Open GeoCMS(Geo-tagged media Contents Management System). Open GeoCMS is a new framework to manage geo-tagged media data on the web. Our framework supports various types which are for moving point, moving photo - a sequence of photos by a drone, moving double and moving video. Also, GeoCMS has the label viewer and editor system for photos and videos. The Open GeoCMS have been developed as an open source system. We will describe the design and implementation of the Open GeoCMS system.

## 2. OPEN GEOCMS ARCHITECTURE

Figure 1 gives the high level architecture of Open GeoCMS system. The purpose of the Open GeoCMS is to provide novel services to store, update, retrieve and manage large amounts of geo-tagged media data. Similar to other contents management systems, our system has been developed and extended to support geo-tagged media data types on a traditional web service framework Apache/Tomcat and PostgreSQL DBMS. Furthermore, one of the most special feature developed in our system is the object detection functionalities associated with TensorFlow framework.

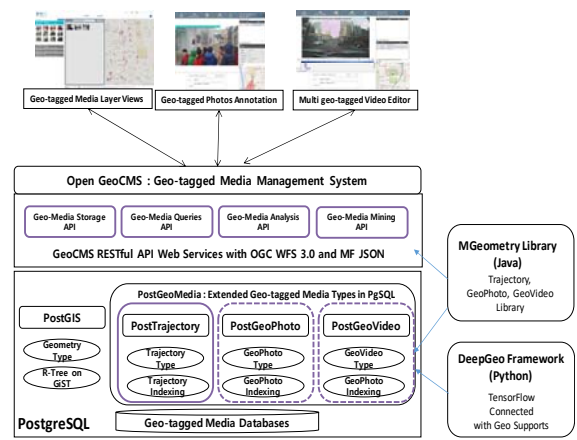

Figure 1. System architecture of Open GeoCMS

---

\* Corresponding author
\+ Both authors contributed equally

Open GeoCMS adopts a layered design of five main layers as described below:

- **PostGeoMedia**: To efficiently store, manage and retrieve geo-tagged media, various user-defined types and functions for geo-tagged media are developed on PostgreSQL DBMS. PostGeoMedia can supports STPhoto, MovingPoint, MovingDouble, MovingPhoto, MovingVideo data types and relationship operators for SQL.
- **GeoCMS RESTful Web Services**: As a geo-tagged API service layer, our RESTful Web Services have been extended on OGC WFS 3.0 draft version which can support GeoJSON. Also, new GeoMedia JSON formats for geo-tagged media has been developed on OGC Moving Feature JSON.
- **MGeometry Library**: Complex topological operators for moving objects and geo-tagged media are supported as a Java library as just like PostGIS make use of C library such as GEOS, Proj.4 and GDAL. We have implemented and tested the MGeometry library together while developing Open GeoCMS.
- **DeepGeo Framework**: Object detection functionalities are supported on various deep learning models which are operated on TensorFlow. The default deep learning model is a MaskRCNN model which are trained by Microsoft COCO dataset. The model can detect 80 object categories including person, car, truck and traffic signs.
- **GeoCMS Applications**: Open GeoCMS suit contains annotation tools and viewer applications for the geo-tagged photos and videos. Users can add text-tags and icons on the photo and video using annotation tools.

In the following sections, we describe the supported data types and the implementation details.

### 3. GEO-TAGGED MEDIA TYPES

Open GeoCMS supports over one dozen data types for geo-tagged data and media to represent and analyse various temporal, spatial, and spatio-temporal media phenomena. In this section, we describe the selected several media types.

### 3.1 Moving Points and Doubles for GPS and GeoSensor

Today, GPS has become very important data for emerging applications such as location-based service and fleet management system. Suppose that a taxi company want to develop fleet management system. They should collect and maintain large quantities of GPS data for tracking taxis. In Open GeoCMS, users can store and manage the GPS trajectories of the taxis using the MovingPoint type.

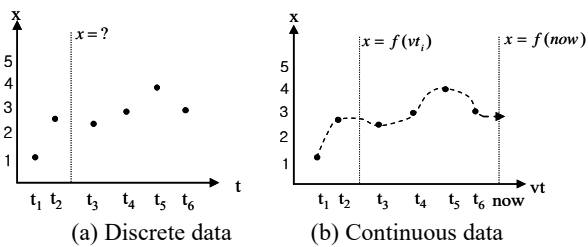

Figure 2. Representation of moving points and values

- **MovingPoint**: MovingPoint is a type to represent locations changes of moving objects as like GPS data of a taxi. When a snapshot of a moving point is a tuple of ( *[latitude, longitude, altitude], time* ), a moving point can be represented as a sequence of the points on the timeline sequence. The following example shows the representation of a moving point in OGC Moving Feature JSON format.

```
{
"type": "MovingPoint",
"cooridniates": [[150.0,50.0,10],[160.0,60.0,12], [170.0, 60.0,11] ],
"datetimes ": ["2018-08-1T13:01:01Z", "2018-08-1T13:01:02Z ",
          "2018-08-1T13:01:03Z"],
"interpolation": "linear"
}
```

As shown in Figure 2(a), the coordinates of a moving taxi are stored on the discrete timeline because GPS cannot be obtained in more than one second. However, user can want to retrieve the coordinates at the 1/2 second between two exact time values. For the interpolation of the time, we can choose the type among the 'discrete', 'linear', 'stepwise', and other arithmetic function.

- **MovingDouble**: MovingDouble is a type for storing geographical sensor values or intermediate after a calculation between moving points. A moving double can be represented as a sequence of double value on the timeline sequence. The following example shows the representation of a moving double for Open GeoCMS.
- 

```
{
"type": "MovingDouble",
"values ": [  5.0, 9.0, 6.0 ],
"timeline": [1533128461000, 1533128462000, 1533128463000],
"interpolation": "stepwise"
},
or
{
"type": "MovingDouble",
"values ": [  5.0, 9.0, 6.0 ],
"timeline": [1533128461000, 1533128462000, 1533128463000],
"coordinates": [ [150.0, 50.0],[160.0, 60.0], [170.0, 60.0]],
"timeline": [1533128461000, 1533128462000, 1533128463000],
"interpolation": "stepwise"
},
```

We extended two keyword sections from the existing OGC Moving Feature JSON. One is the 'timeline' keyword using integer representation. The integer representation can be used in a wide variety of way including user-defined times or UNIX epoch time in milliseconds format. Another is 'coordinate' keyword which can be used for geographically moving sensors. For example, the temperatures can be measured and collected with GPS values while the drone flies through a city.

### 3.2 STPhoto for Geo-tagged Photos

When users share geo-tagged photos or videos, they are provided metadata including position and time, media may also include a title or text label indicating media content. For many geo-tagged media, metadata includes the coordinates of the location where the photo or video was taken.

Furthermore, recent smartphones embedded gyro sensors and magnetometer sensors which can measure the Field of View (FoV) using the information on the camera direction and posture, and can store the information in the EXIF metadata of the image file. Whereas traditional geo-tagged photos only contain the (x, y) information based on GPS in the image file, FoV-tagged photos can include the information such as the direction in which the photo was taken and the angle of the camera at the time of capture. For example, photographing spatial information of spatially-tagged media content captured by a smartphone can serve as basic data for finding spatial regions or points of interest.

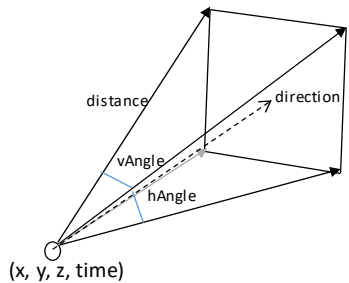

Figure 3. A Field of View in a geo-tagged photo

- **STPhoto :** STPhoto is a type for a geo-tagged photo with a FoV. A STPhoto data is a tuple of *( imguri, loc, t, fov)*. *imguri* is a universal locator of the image file, *loc* is a tuple of ( x, y, h ), and *t* is a timestamp of image which was taken. Also, *fov* is a tuple of *(hAngle, vAngle, direction, distance)* as shown in Figure 3. *hAngle* is the maximum horizontal angle of the camera lens, *vAngle* is the maximum vertical angle of the camera lens, *distance* is the maximum visible distance from the camera, *direction* is the angle from the north to the direction of the camera, and let

```
{
"type": "stphoto",
"uri" : "http://u-gis.net/images/mphoto1.jpg",
"coordinates" : [ -122.0879583, 37.4184889 ],
"timeline" : [ 1533128461000 ],
"fov" : {
    "type": "fov",
    "horizontalAngle": 63,// default: 63 degree for 35mm lens
    "verticalAngle": 60,// default: 60
    "direction2d" : 90, //(optional) North(0), East(90), South(180),
    "distance" : 30 //default:100 m
  }
}
```

The above JSON example shows STPhoto data of a photo which was taken in [-122.0879583, 37.4184889] at timestamp 1533128461000.

### 3.3 MovingVideo for a Geo-tagged Video

Recent advances in mobile, the quantity of geo-tagged video generated by mobile users have significantly increased. While a geo-tagged photo has only one coordinate at a specified time, geo-tagged video can be represented as a sequence of the GPS and FoV information which the video was taken. Also the information changes continuously over time as shown in Figure 4.

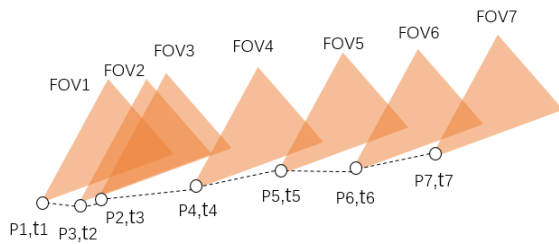

Figure 4. An example of a MovingVideo data

- **MovingVideo:** MovingVideo is a type for a geo-tagged video with a FoV. A MovingVideo data is a tuple of ( *videouri, [locs], [fov], timeline, interpolation)*. A moving video consists of a sequence of frames. In view of FoV a geo-tagged video data can have at least two different type of camera deployment, 1) fixed or straight-ahead direction FoV and 2) the random direction FoV. In fixed or straight-ahead direction FoV, the direction of camera is aligned with the direction of moving object. One of the example is a dashcam which was attached in a vehicle. In random direction FoV, the direction of the camera is randomly varying without an alignment with the direction of moving object which can be generated by a smartphone camera recording. Figure 4 shows an example of the random direction FoV.

```
{
"type": "MovingVideo",
"uri" : "http://u-gis.net/videos/video1.mp4",
"coordinates": [
    [150.0, 50.0],[160.0, 60.0], [170.0, 60.0]
],
"fov" : [  //optional
    {
        "verticalAngle" : 50, //
        "horizontalAngle" : 63,// default: 63 degree for 35mm lense
        "viewDistance" : 30, //default:100m in meters
        "direction2d":90
             //Random :North(0),East(90), South(180), West(270)
             // Fixed: -360(front), -90(right), -180(rear), -270(left)
    },
],
"timeline": [1533128461000, 1533128462000, 1533128463000],
"interpolation": "linear"
}
```

In the next section, we describe the implementation of the Open GeoCMS.

## 4. IMPLEMENTATION OF THE SYSTEM

### 4.1 Portal System for Geo-tagged Media

The portal system of the Open GeoCMS is the web service that deliver geo-tagged media to users from Open GeoCMS data and media repository. Figure 4 shows user interfaces for searching geo-tagged photo and videos.

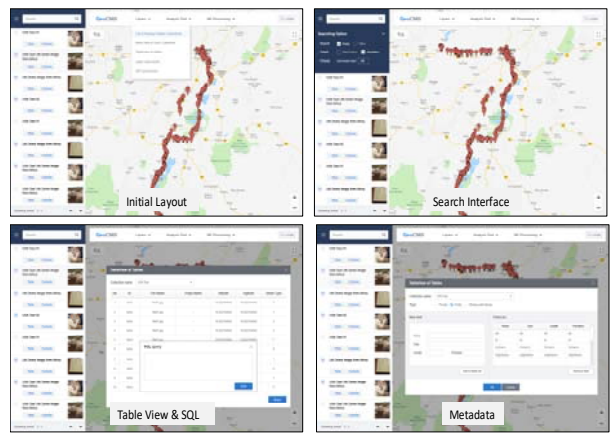

Figure 5. An example of spatial and temporal searching

The portal system of the Open GeoCMS supports the following functionalities for accessing and managing geo-tagged media repository.

- **Geo-tagged Media Layer Management:** Users can create and manage new geo-tagged media layer. Also, user can insert and update new GPS trajectories and photo/videos via the this portal system.
- **Spatial and Temporal Retrieval and Visualization of Geo-Tagged Media**: User can retrieve for and visualize the geo-tagged media using various spatial and temporal predicates.
- **Analysis and Processing of Geo-tagged Media**: User can analyze the geo-tagged media using the similarity and complex queries. Also the media can be transformed and projected into another dimension or data using various processing operators.

This portal system have been developed on Open GeoCMS RESTful API and PostGeoMedia System.

**4.2 Geo-tagged Media Tool and Viewer with Deep Features**

The Open GeoCMS support an authorization tools and viewers for geo-tagged photos and videos. Also, this tools have connected with deep learning functionalities to help automatic detection of geographic and moving objects on the media.

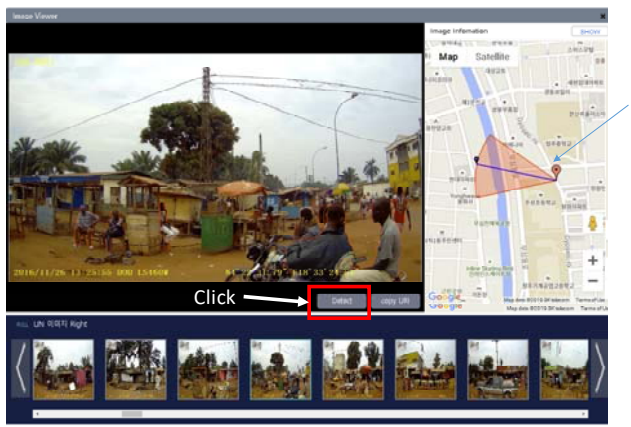

Figure 6. Object detection by deep learning model in a geo-tagged media viewer

The Open GeoCMS includes the following tools that support object detection features by deep learning.

- **Geo-tagged Photo Authorizing Tool and Viewer:** Users can create and check hyper geo-tagged image which has textual, iconic, and polygon-shape annotations by this authorizing tool. Also, the 80 kinds of object classes can automatically detected by MaskRCNN deep learning model, and can be annotated on the image. Figure 6 shows geo-tagged photo viewer and the deep learning annotated objects on it.
- **Geo-tagged Video Authorizing Tool and Viewer:** Users can create geo-tagged video image which has textual, iconic, and polygon-shape annotations by this authorizing tool. As like photo tools, the MaskRCNN deep learning model supports for users to create the annotated geo-tagged video in timeline of the video.

Deep learning features have been developed on TensorFlow framework in python and pl/python. Also, this function was implemented as user-defined functions which can be used in user SQL queries.

**5. CONCLUSION AND FUTURE WORK**

In this paper, we described the design and implementation of the Open GeoCMS system. The Open GeoCMS is a suite of novel and innovative systems that helps users create, manage, and distribute their own geo-tagged media.

To manage and analyse geo-tagged media, a wide variety of high difficulty operators are required. In next version, we would support more various functions to analyse the geo-tagged media.

**ACKNOWLEDGEMENTS**

This research was supported by a Grant (14NSIP-B080144-01) from National Land Space Information Research Program funded by Ministry of Land, Infrastructure and Transport of Korean government. Also part of this research was supported by 2018 free-for-all project from Korea Land and Geospatial Informatix cooperation.

**REFERENCES**

Ding, W., Yang, K. and Nam, K. W.. 2019: "Measuring similarity between geo-tagged videos using largest common view." *Electronics Letters, 55.8, 450-452.*

Ferreira, N., Poco, J., Vo, H. T., Freire, J., & Silva, C. T. 2013: Visual exploration of big spatio-temporal urban data: A study of new york city taxi trips. *IEEE Transactions on Visualization and Computer Graphics*, 19(12), 2149-2158.

Lee, Y., Nam, K. W., Ryu, K. H., 2017: Fast mining of spatial frequent wordset from social database. Spatial Information Research, 25(2), 271-280.

Long, J. A., and Trisalyn A. N.., 2013: A review of quantitative methods for movement data. *International Journal of Geographical Information Science*, 27.2, 292-318.

Kim, K., Ogawa, H., 2017 : OGC Moving Features Encoding Extension – JSON Version 1.0. OGC.